\documentclass[pra, 12 pt]{revtex4}
\usepackage[english]{babel}
\usepackage{amsmath}
\usepackage{epsfig}

\begin{document}
%\draft

\title{
Quantum nature of the critical points of substances
}
\author{S.A. Trigger}

\address{Joint\, Institute\, for\, High\, Temperatures, Russian\, Academy\,
of\, Sciences, 13/19, Izhorskaia Str. 13 bild. 2, Moscow\, 125412, Russia;\\
email:\,satron@mail.ru}

\begin{abstract}

Thermodynamics of chemical elements, based on the two-component electron-nuclear plasma model shows that the critical parameters for the liquid-vapor transition are the quantum values for which the classical limit is absent.     \\

\end{abstract}

\maketitle

For all substances in the vicinity of the critical points of the liquid-vapor transition, the inter-particle interaction is strong. Therefore, theoretical description of critical points seems to be a very complicated problem. Numerical modeling allows for approximate calculation of the critical parameters for a known inter-particle interaction potential. However, the use of pair short-range potentials acceptable for the matter of low and moderate density (in gaseous state) cannot be justified for the region of the critical point parameters. This means that usual numerical calculations of these parameters with the model pair potentials (see, e.g., [1,2]) have empirical nature.

To describe the critical parameters (for concreteness, only the chemical elements are discussed below, although the main statements are universal), we propose to use the pure Coulomb interaction between the electrons and nuclei [3]. This basic model, i.e., the two-component homogeneous and isotropic electron-nuclei Coulomb system (CS) has been recently successfully applied to study the properties of dielectric permittivity [4,5].
It was also shown that the critical point of the two-component CS is related to the limiting behavior of the generalized
screening length in the electron-nuclear plasma [6].

In this Letter, we focus attention that in two-component electron-nuclear Coulomb plasma, when the thermodynamic parameters tend to critical, the sole parameter with energy dimension (excepting the parameters containing the critical ones) is the atomic energy unit $me^4/\hbar^2=2Ry=27,21$ eV. It immediately follows from this statement that the critical temperature $T_c$ (in energy units) is given by

\begin{eqnarray}
T_c= \frac {m e^4}{\hbar^2}\, \tau\left(z, \frac{m}{M}\right). \label{F1}
\end{eqnarray}
Here $m$, $e$ are the electron mass and charge, respectively, $M$ and $ze$ are the mass and charge of nuclei of the element under consideration. The function $\tau (z,\frac{m}{M})$ is the unknown dimensionless function of two dimensionless parameters $z$ (nuclear charge number) and the mass ratio $m/M$. From physical reasons, we can assume that the dependence of the critical temperature on the small parameter $m/M$ can be neglected with good accuracy. Then the problem reduces to the determination of only the function $\tau (z)$

\begin{eqnarray}
T_c \simeq \frac {m e^4}{\hbar^2}\tau(z). \label{F2}
\end{eqnarray}

For the critical pressure, on the same basis as above one can write
\begin{eqnarray}
P_c = \frac {m e^4}{a_0^3\hbar^2}\, \pi\left(z, \frac{m}{M}\right)\simeq \frac {m e^4}{a_0^3\hbar^2}\, \pi(z), \label{F3}
\end{eqnarray}
where $a_0=\hbar^2/me^2$ is the Bohr radius.
At last, the critical pressure can be written in the form
\begin{eqnarray}
N_c = \frac {1}{a_0^3}\, n \left(z, \frac{m}{M}\right)\simeq \frac {1}{a_0^3}\, n(z). \label{F4}
\end{eqnarray}
The introduced dimensionless functions $\pi$ and $n$ under condition $m/M \ll 1$ depend only on the nucleus charge number $z$.

Consideration of the experimental data for the critical points for the liquid-vapor transition shows that the function $\tau (z)$ is a rapidly varying function of the variable $z$. One can assume that these rapid "irregular" changes in the function $\tau (z)$ have the same physical nature as spasmodic changes in the first ionization potential $I(z)$ and valence $v(z)$ which are conditioned by sequential filling of electron shells in the atomic model of matter. However, it should be emphasized that the ionization potential is an approximate model characteristic related to the rarified (gaseous) state of matter. Therefore, the above assumption has a limited and very model nature. At the same time, the representations (\ref{F1})-(\ref{F3}) are exact, although, in such a general form they cannot provide prediction of the critical points of some elements on the basis of the known critical points of other elements. For such predictions, methods for calculating or physical models of the functions $\tau(z)$, $\pi(z)$ è $n(z)$ should be developed.

Useful information can be obtained already within the existing model approximations. As an example, let us consider the Van-der-Waals theory. As is known the relation
\begin{eqnarray}
P_c = \frac {3 N_c T_c}{8}, \label{F5}
\end{eqnarray}
follows from this theory.

Although relation (\ref{F5}) is found from the classical and empirical approach, it can be used within the range of its practical applicability to obtain the approximate relation between the functions  $\tau(z)$, $\pi(z)$ è $n(z)$. The Van-der-Waals model leads to the relation

\begin{eqnarray}
\pi(z)= \frac {3}{8}\, n(z)\tau(z). \label{F6}
\end{eqnarray}

Similarly, the modern model theories of the equation of state and critical points of matter (see, e.g., [7] and references therein) can be used to approximate the unknown functions $\tau(z)$, $\pi(z)$ è $n(z)$ according to the known experimental data and the periodic Mendeleev Table of chemical elements. However, it is not the problem of this Letter.

As follows from the basic relations  (\ref{F1})-(\ref{F3}) \emph{the parameters of the critical points are the quantum expressions which have no a classical limit or classical analogue}. In this connection, the general question, outside the frameworks of this Letter, arises about the classification of the physical characteristics of the Coulomb matter into those having a classical limit or not.

\section*{Acknowledgments}

I am thankful to V.B. Bobrov for the useful discussions.

This study was supported by the Netherlands Organization for
Scientific Research (NWO), grant no. 047.017.2006.007 and the Russian Foundation for Basic Research, project no. 07-02-01464-a.

\end{document}